\documentclass[]{spie}  
\usepackage[]{graphicx}

\title{PIONIER: a visitor instrument for the VLTI} 

\author{ J.-P., Berger\supit{a,b} , G. Zins\supit{a} ,
  B. Lazareff\supit{a} , J.-B. Lebouquin\supit{a} ,
  L. Jocou\supit{a} , P. Kern\supit{a} , R. Millan-Gabet\supit{c} ,
  W. Traub\supit{d} , P. Haguenauer\supit{b} , O. Absil\supit{e} ,
  J.-C. Augereau\supit{a} , M. Benisty\supit{f} ,
  N. Blind\supit{a} , X. Bonfils\supit{a} ,
  A. Delboulbe\supit{a} , P. Feautrier\supit{a} ,
  M. Germain\supit{a} , D. Gillier\supit{a} , P. Gitton\supit{b}
  , M. Kiekebusch\supit{b} , J. Knudstrup\supit{b} , J.-L
  Lizon\supit{b} ,
  Y. Magnard\supit{a} , F. Malbet\supit{a} , D. Maurel\supit{a}
  , F. Menard\supit{a} , M. Micallef\supit{a} ,
  L. Michaud\supit{a} , S. Morel\supit{b} , T. Moulin\supit{a} ,
  D. Popovic\supit{b} , K. Perraut\supit{a} , P. Rabou\supit{a} , S. Rochat\supit{a} ,
  F. Roussel\supit{a} , A. Roux\supit{a} , E. Stadler\supit{a} and
  E. Tatulli\supit{a}
\skiplinehalf
\supit{a}LAOG,UJF,CNRS, 414 Rue de la Piscine, 38400 Saint Martin
d'Heres , France; \\
\supit{b} European Southern Observatory, Paranal, Chile, Garching, Germany\\
\supit{c} NextSci, California Institute of Technology, Pasadena, California USA\\
\supit{d} Jet Propulsion Laboratory, California Institute of
Technologie, Pasadena, California, USA \\
\supit{e} Universite de Li\`ege, Li\`ege, Belgium \\
\supit{f} Osservatorio Astrofisico di Arcetri,INAF,  Arcetri, Italy
}

\authorinfo{Further author information: \\J.-P. Berger:
  jpberger@eso.org\\  J.-B. Le Bouquin:
  Jean-Baptiste.Lebouquin@obs.ujf-grenoble.fr \\
G. Zins: Gerard.Zins@obs.ujf-grenoble.fr
}

\begin{document} 
\maketitle 

\begin{abstract}
  PIONIER is a 4-telescope visitor instrument for the VLTI, planned to
  see its first fringes in 2010. It combines four ATs or four UTs
  using a pairwise ABCD integrated optics combiner that can also be
  used in scanning mode. It provides
  low spectral resolution in H and K band. PIONIER is designed for
  imaging with a specific emphasis on fast fringe recording to allow
  closure-phases and visibilities to be precisely measured. In
  this work we provide the detailed description of the instrument and
  present its updated status.
\end{abstract}



\section{Introduction}
\label{sec:intro}  

Long baseline interferometry is the only tool capable to provide
spatially resolved information with milli-arcsecond resolution.  Over
the last two decades an increasing number of astrophysical topics has
taken benefit of such interferometric observations in the visible and
the infrared. Interferometers such as COAST, NPOI,
IOTA and PTI have paved the way for larger facilities. The old
visibility fitting paradigm has evolved and the richness of the
interferometric observables has lead to new powerful
diagnostics. Today, second generation interferometers such as VLTI,
CHARA and KeckI provide spectrally dispersed visibilities, phases and
closure phases. The ultimate goal for an optical interferometer, to
achieve aperture synthesis imaging like in the radio domain, is now a
reality.\smallskip

Until very recently, image reconstructions were limited to multiple
systems\cite{Baldwin:1996,Benson:1997,Monnier:2004a} with no extra
information added by the image. It is only very recently that
J. Monnier and his team obtained the first scientific meaningful
images using the MIRC instrument and four CHARA telescope on Mt
Wilson. The spectacular images of the surface of
Altair\cite{Monnier:2007} or the interacting binary $\beta$ Lyrae
\cite{Zhao:2008a} contained unexpected features. More recently
Lebouquin et al.\cite{LeBouquin:2009} imaged the surface of the MIRA
star T Lep using AMBER at the VLTI while Kloppenborg et
al.\cite{kloppenborg:2010} captured the dramatic eclipse of Epsilon
Aurigae by an accretion disk. These results are the demonstration that
aperture synthesis in the optical domain is operational and that
reconstructed images bring a real added value to the standard Fourier
domain complex visibility fitting.

VLTI second generation instruments Gravity (Gillessen et al,
these proceedings) and Matisse (Lopez et al. these proceedings) will provide
the astronomical community with an imaging capability respectively in
the near-infrared and mid-infrared. But these instruments will not be
operational prior to 2014.

PIONIER is a visitor instrument designed to provide VLTI, by the end
of 2010, with a new observational capability that combines
imaging and precision. It is operated either in H or K band with two
spectral resolutions: broadband and $R\sim40$ (eventually $\sim 100$). It will be complementary to the
current near-infrared AMBER instrument\footnote{AMBER provides VLTI
  with high spectral resolution capability in H and K.}. The interferometric
combination relies on the integrated optics (IO) technology to combine
four VLTI beams (Auxiliary Telescopes or Unit Telescopes). It
provides, in one observation, the simultaneous measurements of 6
visibilities and 3 + 1 closure phases. The paper is organized as
follows: section \ref{sec:science} discusses PIONIER science
cases and section \ref{sec:description} describes the instrumental
concept. Section \ref{sec:data} is devoted to the description of the
interferometric signal and its reduction and we conclude in section
\ref{sec:status} with the current status of the
instrument and its future developments.

\section{Science cases for PIONIER}
\label{sec:science}

 \begin{figure}[t]
   \centering
   \includegraphics[width=0.4\textwidth]{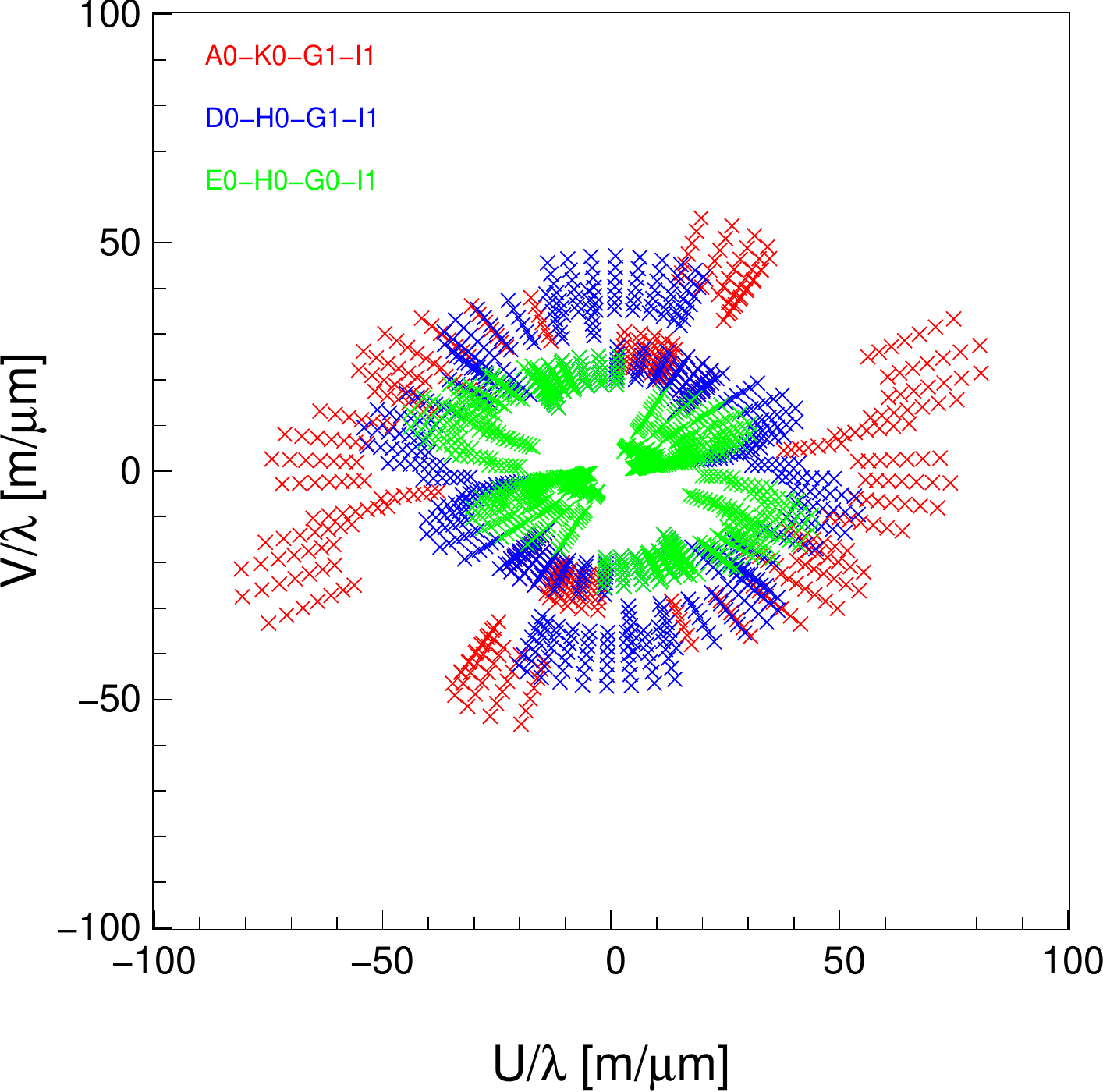} 
 \caption{Typical {\it uv} coverage on a southern source with PIONIER and 3
   quadruplets configuration and a spectral resolution of 100.}
   \label{fig:uv}
 \end{figure}

PIONIER has been designed to provide a low spectral resolution imaging
capability to the VLTI and to push precision interferometry has far as
possible. Figure \ref{fig:uv} shows an example of {\it uv} coverage
obtained using three AT quadruplets.

Although many astrophysical topics will
eventually benefit from its characteristics we have a core program that is related to star
formation and exoplanetary studies (exozodi and planets).

\paragraph{Morphology of
  protoplanetary environments around intermediate mass pre-main sequence stars}

Herbig Ae Be stars, the intermediate-mass analogs of T Tauri stars
have been intensively studied with optical interferometers in the last
13 years (see e.g Millan-Gabet et
al.\cite{MillanGabet:2007}). Consequently the view on the inner
structure of accretion disks has been substantially revisited to take
into account such milli-arcsecond resolutions direct observations.
The concept of ``puffed-up'' inner rim has been introduced. It marks
the boundary in the disk where dust can no longer exist due to
sublimation and forms a barrier to the photons causing altering the
vertical structure of the disk. Additionally there are several
evidences that matter flows through this limit to the star and the
``hard'' limit represented by the dust inner rim is now investigated
with care. Revealing the nature of the matter inside the dust
sublimation limit and the exact smoothness of the transition is still
an active area of research.  In order to contribute to that study we
have identified a small sample of HAeBes whose environment have been
clearly resolved by the VLTI and should therefore be ideal candidates
for imaging with PIONIER in order to reveal the morphology and
time-variability of the inner astronomical unit near-infrared
emission.

\paragraph{The shape of T Tauri disks}

If sensitivity allows it PIONIER will be used to probe
the structure of the inner dust disk around T Tauri stars. While the
circumstellar disks are not expected to be fully resolved the
visibilities will give us access to the inner disk location while the
closure phases will allow the morphology and surface brightness
profile to be probed. In the best cases, it will be possible to
extract the inner rim position, ellipticity and thickness. All of
these are of major importance to understand the conditions for star
and planet formation. 

\paragraph{Origin, morphology and evolution of hot inner debris discs
  around main sequence stars}

The inner Solar System contains a cloud of small dust grains created
when small bodies – asteroids and comets – collide and outgas. This
dust cloud (the zodiacal disc) has long been suspected to have
extrasolar analogs (exozodiacal discs), which have remained elusive
until recently, when near-infrared interferometric observations
revealed small (∼1\%) excesses around several nearby stars \cite{Absil:2006,Absil:2008}. These excesses have been
interpreted as being due to hot, possibly transient exozodiacal dust
within 1 AU from these stars. We envision to use the high-precision
visitor instrument PIONIER in order to (i) characterise the hot dust
grains properties, (ii) tentatively study the inner disc morphology
in bright cases, and (iii) survey young main sequence stars to study
the hot dust content during the late phases of terrestrial planet
formation. Given the importance of precision such observations will be
preceded by an adequate commissioning program of PIONIER accuracy in
order to determine its performances and proceed (or not) with the
science program.

\paragraph{Direct detection of Hot Jupiters with infrared
  interferometry}

Closure phase is one of the most robust interferometric
observables. By construction it suppresses the harmful effect of
atmospheric turbulence. Zhao et al.\cite{Zhao:2008b} have proposed and
experimented the use of closure phase measurements at CHARA to spatially
resolve Hot Jupiters (HJ) which elude direct detection from single-mirror
instruments. {\bf In order to achieve such a result a great deal of
systematics biases has to be hunted down and understood which is far
from easy}. PIONIER has
such an experimental program which will start with a dedicated
commissioning. The success of the preliminary testing campaing will
determine whether the program will continue.
If succesful PIONIER will have the means to separate the flux contribution of the
planet from that of its parent star and potentially obtain low
spectral resolution information. From low-resolution spectra of
HJs obtained with PIONIER, model fitting shall: (i) provide a
measurement of their albedo and temperature, (ii) test the cloud-free
assumption, (iii) constrain the heat redistribution, and (iv) detect
CO and H2O species. Such major observations of the still poorly known
HJs will obviously put unprecedented constraints on our understanding
of planetary structure and formation.

\paragraph{Other studies}

The previous science programs bear with them some risk which is linked
either to the intrinsic magnitude of the sources or to the
precision requested on the interferometric observables
measurements. We have therefore a bright object program centered
imaging of stellar surfaces and the study of interacting
binaries. However, a number of other programs such as e.g Wolf-Rayet studies,
or low-mass stars diameters studies etc. will be developed in the context of
external collaborations.

\section{PIONIER Description}
\label{sec:description}

PIONIER is a 4 beam combining instrument that  will be located on the VINCI
table in the VLTI laboratory. It will be operated initially in the H
band and later in the K band. The science cases of PIONIER have driven
some two high level technical specifications. The request for precise
visibility measurements was translated into a fast fringe
recording capability in order to beat or calibrate for adverse effects such as atmospheric piston,
photometric coupling fluctuations and vibrations. The request for
sensitivity lead us to define a ``broadband'' mode (without spectral
dispersion) and a particular care into the output imaging optics.

The following subsections give a global
overview and go into some detail of some of the instrument functions.

\subsection{Subsystem description, layout and optomechanical structure}

\begin{figure}[p]
  \centering
  \includegraphics[width=0.7\textwidth]{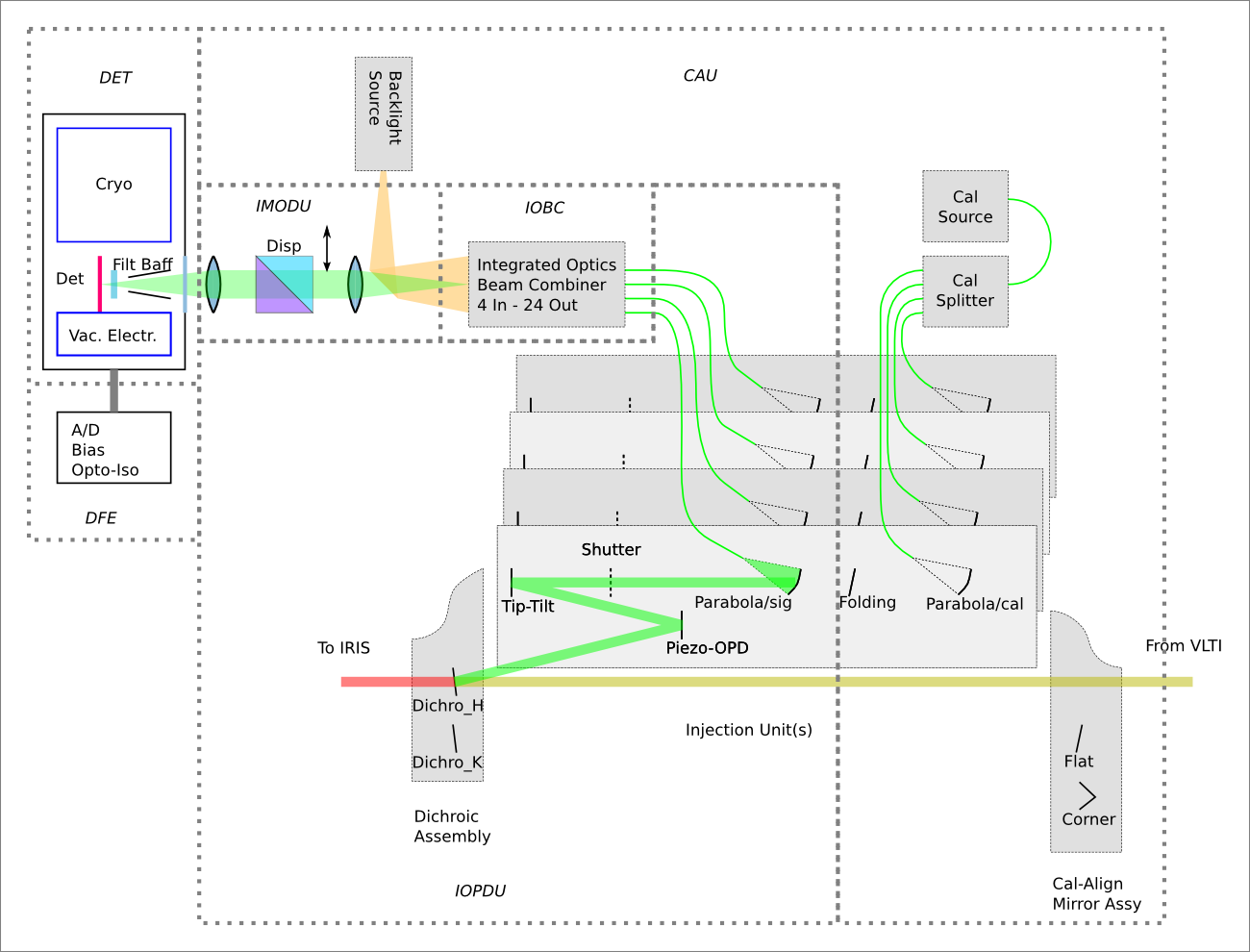}
  \caption{Pionier conceptual scheme}
  \label{fig:concept}
\end{figure}

\begin{figure}[p]
  \centering
  \includegraphics[width=0.7\textwidth]{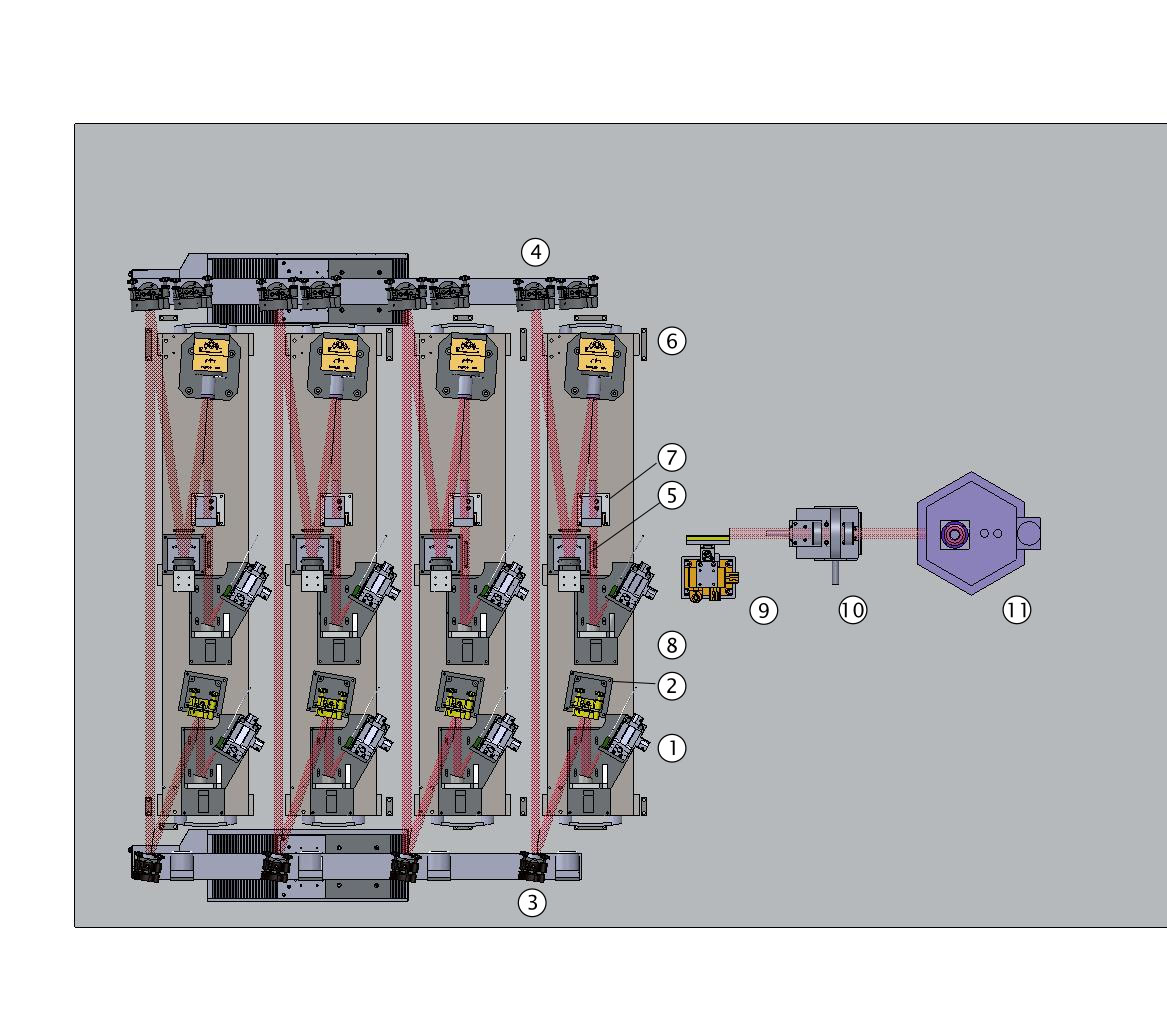}
  \caption{Pionier optomechanical and implementation overview}
  \label{fig:optomecha}
\end{figure}

The optomechanical layout is shown on Figure~\ref{fig:concept} Functionally, it
consists of the following subsystems:
\begin{itemize}
\item Injection and Optical Delay Unit (IOPDU). Its function is to
  inject the free-space beams from the VLTI into optical fibers. 
It includes a tip-tilt correction to optimize the fiber coupling and
and a piezo-electric scanning translation to allow for optical delay
sweeps of up to 800$\mu$ m amplitude.
\item Calibration and alignment Unit (CAU). Its serves dual function:
  (i) to inject, as a substitute for the astronomical beams, mutually
  coherent beams derived from a common Tungsten source, for laboratory
  verification and health check purposes; (ii) to reflect, via a
  corner cube mirror, a reverse-propagating beam from the output of
  the IOBC (see below) towards the IRIS beam tracker of the VLTI
  observatory, for the purpose of verifying the alignment of the
  Pionier instrument with the IRIS reference positions.
\item Integrated Optics Beam Combiner (IOBC). It takes as input,
  through optical fibers, the signals from the four VLTI telescopes,
  and delivers 6x4 outputs, consisting, for each of the six telescope
  pairs, of their addition with 0, $\pi$/4; $\pi$/2, and 3$\pi$/4 phase lags.
\item Imaging Optics and Dispersion Unit (IMODU). It images the 24
  outputs of the IOBC onto the camera's focal plane, with an
  intermediate space where the image is at infinity, allowing the
  insertion of a dispersing prism and/or a Wollaston polarization
  diplexer.
\item Camera (DET), comprising the cryostat, detector and internal electronics.
\item  More electronic functions --including the digitizing of the
  video signal-- are provided by the Detector Frontend Electronics
  (DFE).
\item Control system (CS) that includes hardware control of the instrument
  units, detector readout, quicklook and interaction with VLTI.
\end{itemize}

One sliding arm allows PIONIER to be set into three configurations:
\begin{enumerate}
\item free: let VLTI beams reach the instrument and operate in
  ``science mode"; 
\item mirror: send coherent signal to the instrument for internal
  calibration purposes; 
\item corner-cube: image the input fibers onto IRIS for alignment check.
\end{enumerate}

Figure \ref{fig:optomecha} shows the instrument in calibration configuration, with the
modules numbered as follows:
\begin{enumerate}
\item Fiber optics to beam injection (via an offset parabola) for the calibration signal;
\item Folding mirror;
\item Second folding mirror; 
\item Dichroic mirror. When observing the sky signals in either H or K
  band, the other band is transmitted to the IRIS camera for real-time
  monitoring of beam offsets.
\item Optical path difference modulator, consisting of a flat mirror
  on a piezo translation stage, with 400um mechanical travel (800um
  OPD).
\item Tip-tilt corrector. Also based on piezo actuators, it receives
  correction inputs from IRIS. Note that the correction is open-loop.
\item Shutter for obtaining dark exposures.
\end{enumerate}

\subsection{Detector}

The detector is a PICNIC camera previously used at PTI (JPL/Caltech)
and IOTA (Harvard- SAO Center for Astrophysics). Its electronics will
be partially upgraded to improve the speed efficiency (see Pedretti et
al.\cite{Pedretti:2004}). One of the challenges of PIONIER will be to
overcome rapid piston and coupling fluctuations and, on the UTs, to
overcome the vibrations. This explains why a specific effort is put
into sampling as fast as possible the detector pixels.
PIONIER detector will have two pixel distributions to read:
\begin{itemize}
\item {\bf broadband mode:} only one row of 24 pixels is read, this should bring
  sensitivity and, because of increased speed, precision;
\item {\bf spectrally dispersed mode}: 24$\times$7 (or $\times$3) pixels have to be
  read. This will allow the {\it uv} coverage to be increased and
  should therefore be used for imaging.
\end{itemize}

\subsection{Injection and OPD Control Unit}

PIONIER's tip/tilt correction comes in addition to the one provided by the
Auxiliary Telescopes (ATs) or the Unit Telescopes (UTs) and allows
the additional tip/tilt contributions coming from the tunnel
turbulence to be compensated for. The VLTI IRIS infrared array
provides the beam angle of arrival measurements.

The optical path difference control unit has a long stroke (800
$\mu{m}$) that
allows long opd drifts to be corrected and relaxes the need for
high frequency delay line control.

\subsection{Calibration and Alignment Unit}
The CAU is designed to provide: 
\begin{itemize}
\item Four beams aligned with
IRIS/MARCEL that can serve as an internal alignment reference; 
\item Four
co-phased beams that allow to record internal fringes purposes. 
\end{itemize}

The CAU is made of: 
\begin{itemize}
\item Two sources: halogen lamp and 1.55 $\mu$m laser; 
\item An integrated optics 1x4 splitter; 
\item Four collimators; 
\item Four backwards reflecting mirrors. 
\end{itemize}
The beams provided by CAU are retractable through
the use a mirror + corner-cube motorized unit with precise
positioning.

\subsection{The integrated optics beam combiner}

\begin{figure}[t]
  \centering
\begin{tabular}{cc}
  \includegraphics[width=0.6\textwidth]{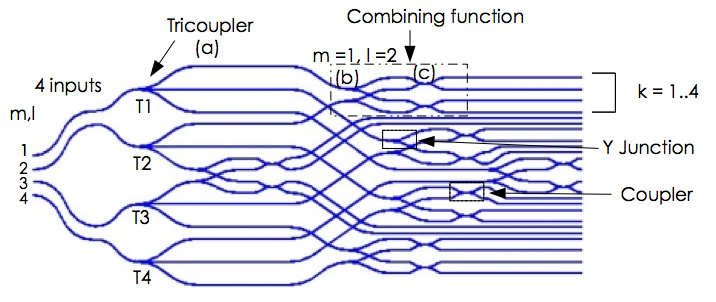}
  &   \includegraphics[width=0.4\textwidth]{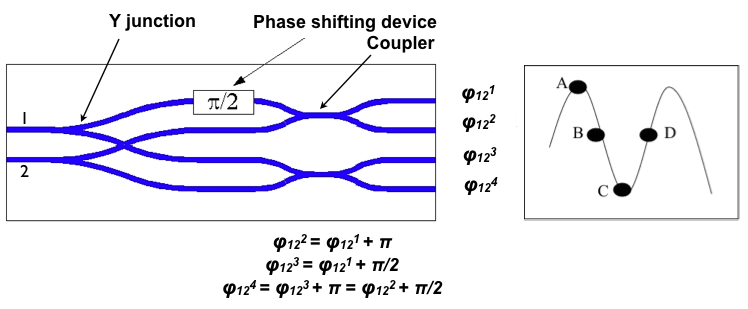} \\
\end{tabular}
  \includegraphics[width=0.4\textwidth]{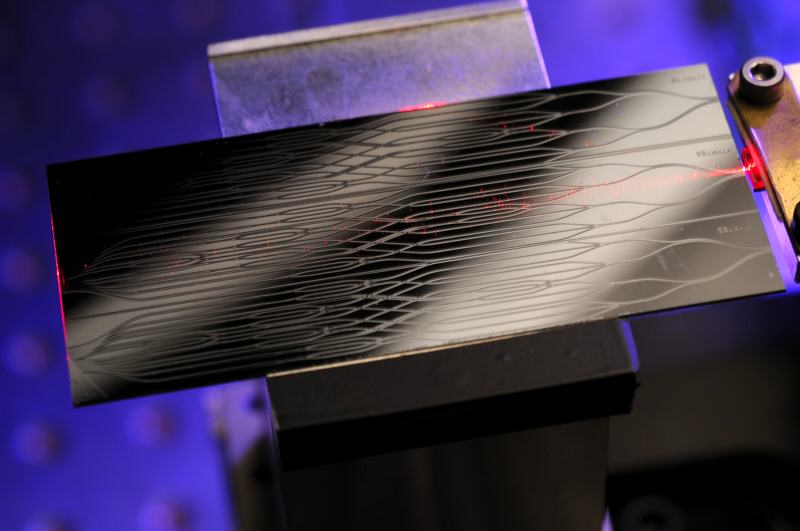}
  \caption{Top left: 4T Beam combination scheme. Top right: detail on
    the 2T ABCD beam combination function. Bottom: The picture of the
    IO combiner without is fibers. Three combination circuits can be
    seen on the figure, only one is uded.}
  \label{fig:combination}
\end{figure}

The integrated optics (IO) combiner is described in detail in Benisty
et al.\cite{Benisty:2009} and can be seen in bottom of figure
\ref{fig:combination}. The four incoming beams (top left part of
figure \ref{fig:combination}) are split in three and distributed in
the circuit to allow a pairwise combination to be done. A so-called
“static-ABCD” combining cell is implemented for each baseline. It
generates simultaneously four phase states (almost in quadrature) that are used to
extract the coherent complex visibility (top-right of figure
\ref{fig:combination}). The  low-chromaticity phase shift is obtained through the use of
specific waveguides with carefully controlled refraction
index. Consequently: 24 outputs have to be read. This
combiner can be used both in fringe-scanning mode (Vinci-like) or
ABCD-like mode.  There is currently one combiner available in the H
band and one that is being developed in the K band.

Since we are using birefringent fibers and the IO is intrinsically birefringent
differential propagation can lead to dramatic contrast reduction. The
final polarization control strategy is not determined yet and several
solutions are contemplated. The two favorites ones are polarization
splitting using a Wollaston (the IOTA way) or active polarization
compensation using free space birefringent plates. Both of them are
being tested in laboratory.

\subsection{Imaging Optics and Dispersion Unit}

\label{sec:ass-imaging}

The first generation imaging and dispersing unit is the one that was
used at IOTA. It images the 24 aligned outputs of the IOBC onto the
detector array with or without spectral dispersion (3 or 7 spectral
channels across the H band) in the perpendicular direction. Without
dispersion, each output is imaged on a single pixel of the detector. A
second generation optical relay unit that should accomodate the H and K
band combiner is under study.

The optical layout is presented figure \ref{fig:Opt-Layout} and has the following specifications:

\begin{figure}[p]
\centering
\includegraphics[width=0.7\textwidth]{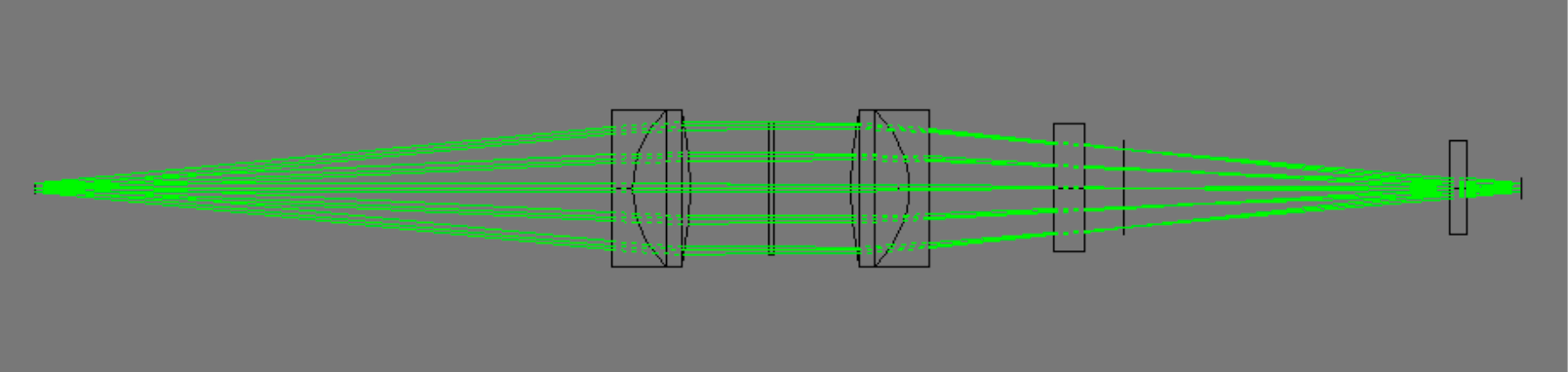}
\caption[Design of the imaging system.]{Optical layout of the imaging
  and without dispersor. Double triplet + detector window + filter
  wheel to scale.}
\label{fig:Opt-Layout}
\end{figure}

\begin{figure}[p]
\centering
\includegraphics[width=0.7\textwidth]{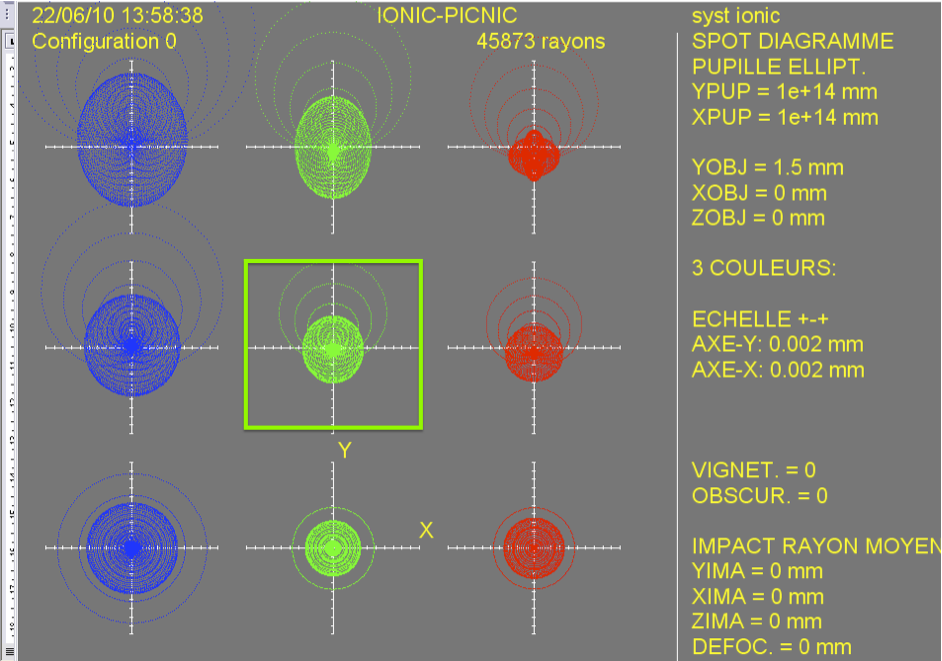}
\caption[Image quality of the imaging system]{Spot diagram of the
  imaging system (from left to right $\lambda = 1.45,1.65,1.8
  \mu\rm{m}$). The pixel size (40 $\mu \rm{m} $) appears as a square box.}
\label{fig:Opt-Quality}
\end{figure}

\begin{figure}[p]
  \centering
  \includegraphics[width=0.4\textwidth]{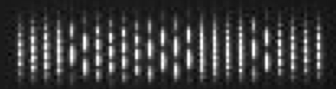}
  \caption{Laboratory image of the dispersed output of the combiner.}
  \label{fig:imaging}
\end{figure}

\begin{itemize}
\item System optics length of $180 mm$ 
\item Focal length of $110 mm$
\item System optics clear aperture of $27mm $
\item Collecting aperture of F/5 
\item Magnification of 1
\item Image quality (see Fig. \ref{fig:Opt-Quality}): The whole flux for each output is focused in the PICNIC camera pixel (Pixel size: $40\mu m$).
\item Image field of view of 3 mm corresponding to the output geometry (24 outputs with $160 \mu m$ spacing) imaged on the detector with 3 pixels between each output image.
\item System dispersion based on a zero-deviation prism to simplify the opto-mechanical implementation. It is composed of standard infrared glasses. The dispersion provided by the prism is spread out the signal over three or seven pixels across the H band according to the chosen prism.
\end{itemize}

\subsection{Control and software architecture}

\begin{figure}[t]
  \centering
  \includegraphics[width=0.6\textwidth]{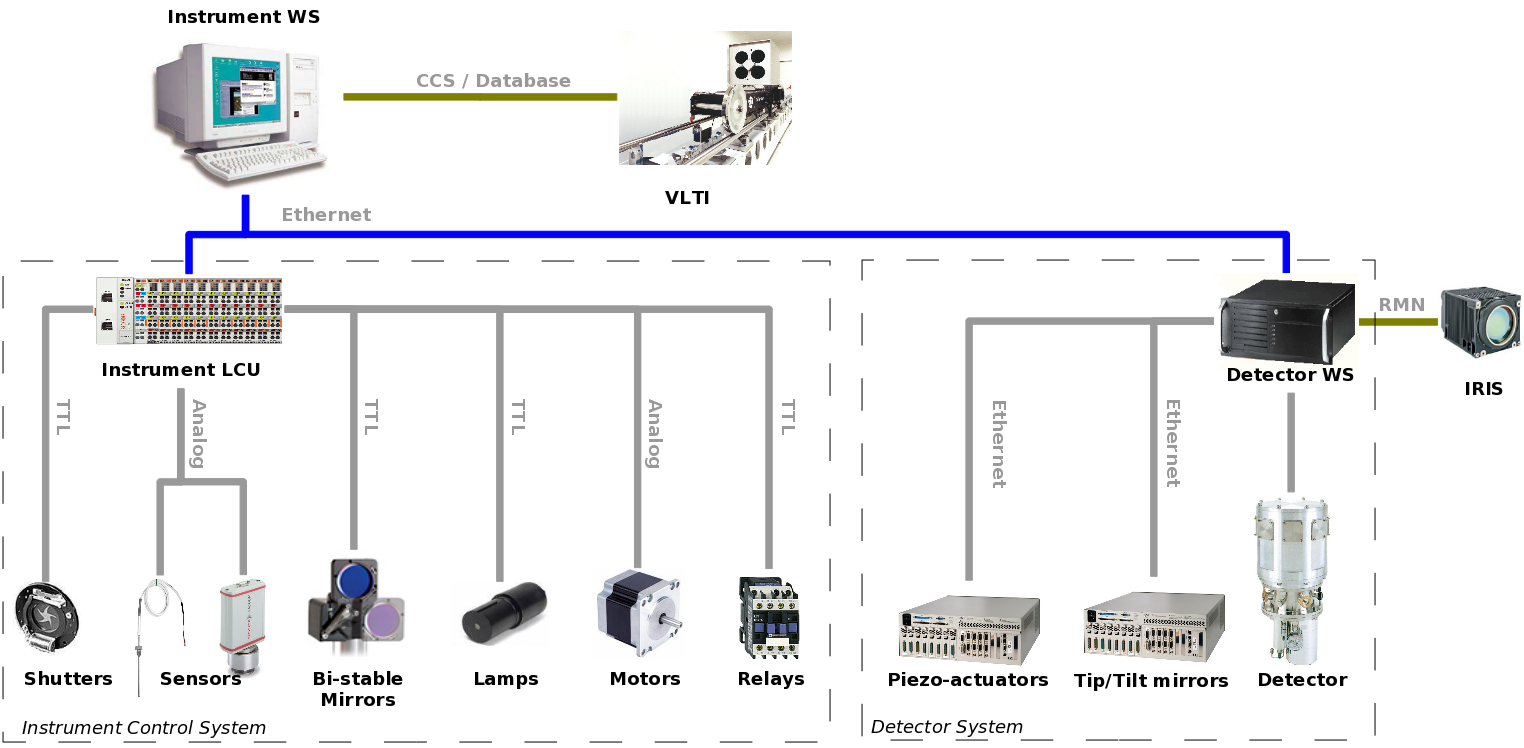}
  \caption{Schematics of PIONIER hardware/software control architecture.}
  \label{fig:control}
\end{figure}

The hardware architecture has been split into 2 parts; detector system
and instrument control system. It is similar to a standard VLTI
instrument with an instrument workstation (WS), an Instrument Control
System (ICS) Local Control Unit (LCU) and a Detector Control System
(DCS) WS, with dedicated tip/tilt and OPD correction facilities. The
instrument WS is a standard Linux PC platform supported by the VLT
Software. The detector WS is an industrial Linux PC platform equipped
with a PCI-7300A board used to generate clock signals and to acquire
detector data. It is running under the VLT Software. The instrument
LCU is an Embedded PC from BECKHOFF New Automation Technology with
EtherCAT/fieldbus terminals for I/O.

PIONIER electronics will be housed into a cooled cabinet close the
instrument optical table. This will include the controller detector
front-end electronics, the DC detector power supplies together with
the detector WS, tip/tilt and scanning piezo controllers, EtherCAT,
calibration lamps and control electronics for electro-mechanical
devices.

The Instrument Software Package
is subdivided into the following standard INS software modules: 
\begin{itemize}
\item[ICS] The Instrument Control Software controls all devices which
  belong to the instrument except the detector. It is based on
  standard software provided with VLT SW, called Base ICS (ICB). 
  ICB has been extended to support Ethernet base fieldbus systems as used
  for PIONIER.
\item[DCS] The Detector Control Software carries out all tasks to
  control the detector subsystem, to perform real-time image processing
  and when needed transfer detector data to the workstation. It is based
  on ESO NGC's software that has been adapted to handle the PICNIC. It
  provides also a quicklook capability to display interferograms and
  first observables raw estimation (visibility, closure phase).
\item[OS] The Observation Software is used to coordinate the execution
  of an exposure for a given observing mode. It provides for setup and
  coordination of the various control systems such as instrument,
  detector, telescope and also interfaces to other software like the
  Archive system to archive observation data. It also completes the
  final FITS header for the observation data file.  The OS does not
  access hardware functions of the instrument. It has the ‘knowledge’
  of how to coordinate the control systems to perform exposures for
  given observing modes. It is based on Base Observation Software Stub
  (BOSS) provided with VLT SW.
\item[MS] The Maintenance software consists in technical templates, which are used for instrument configuration, check-out and troubleshooting.
\end{itemize}

Observing templates are executed using the standard ESO BOB
tool. Dialog between PIONIER and VLTI (e.g delay lines) is achieved
through the VLTI Interferometer Supervisor Software

\subsection{Interface with VLTI}

PIONIER is located inside the VLTI lab on the table of the late VINCI
instrument. PIONIER intercepting optics can be automatically removed
at any time to allow to exploit the VLTI instruments.  A new automatic
filling unit has been designed to accomodate as much as possible VLTI
laboratory specific requests.
Telescopes and delay lines are preset using the standard VLTI
software. PIONIER workstation dialogs with VLTI through ISS to off-load
low-frequency OPD corrections and through RMN to get centroid
information from IRIS detector.

\subsection{Operating modes}
Prior to the scientific operations, the following calibrations are
expected to take place on a daily basis on the internal sources of
VLTI.
\begin{itemize}
\item The internal flux splitting ratio, internal contrasts and
  internal phase shifts can be measured by means of long-scan
  fringes obtained with the internal piezo-electric devices and with
  the 4T coherent source MARCEL from VLTI.
\item The relative alignment between the VLTI axes and the PIONIER
  axes can be measured in the late afternoon with the coude beacon on
  the Auxiliary Telescopes. The measured offsets are stored and serve
  as a reference for the night.
\end{itemize}
Afterward, a typical PIONIER observation is based on the following
sequence:
\begin{itemize}
\item VLTI preset: The new coordinates are sent to the VLTI and its
  subsystem. The preset time is expected to take between 2 to 5min
  depending on the distance and the brightness of the target. Few
  times in the night, the preset also includes the check of the VLTI
  pupil position, when the observed star is brighter than V=4. It
  requires about more 5min. At the end of the preset, the star is
  aligned and guided on the reference axis of VLTI and light should
  enter PIONIER. The VLTI Delay Lines are in blind trajectory.
\item Flux optimization: The internal tip-tilt of PIONIER are used to
  maximize the amount of flux in the PICNIC camera. A small grid of
  offset positions are stored and fit by a Gaussian. The duration is
  about 5min. The relative stability between PIONIER and the VLTI axes
  is expected to be good. We anticipate that this optimization is
  required only few times during the night (if any).
\item Fringe search: OPD offsets are sent to the VLTI delay lines
  until the fringes are seen in the PIONIER real-time display. As of
  now, this search is manual in order to minimize the interface
  between PIONIER and VLTI. According to the accuracy of the VLTI
  pointing models, we expect this search to take between 30s to 10min
  depending on the difficulty of the target (first time being
  observed, faint, resolved, long baselines).
\item Fringe recording: Once the Delay Lines offsets have been found,
  fringes integration is started. The fringes are temporally scanned
  across the coherence packet by the beam of long-range piezo-electric
  devices (although a non-scanning mode can be implemented latter on,
  using the simultaneous ABCD outputs). A typical exposure file
  contains about 200 scans. PIONIER is designed to ensure its own
  group delay tracking. At the end of each scan, a group delay is
  estimated and used in a servo loop to correct the zero positions of
  the scanning piezo-electric devices. The average group-delay is
  forwarded to the VLTI delay lines on a very slow rate (once every
  minute).
\item Internal flux splitting ratio and dark recording: A quick
  sequence of shutter (one beam at a time, then all beam closed) is
  used to estimate the internal flux splitting ratio and the dark
  level of the PICNIC camera. According to the stability obtained in
  commissioning, this step may be skipped or not.
\end{itemize}

PIONIER performs observations in two modes: broad-band and dispersed
($R \approx 35$). The switch between these modes requires to manually
insert/remove the disperser prism in the collimated part of the beam
after the beam combiner, see Figure~1. The broad-band mode will
obviously provide the best sensitivity. It may also provide the best
precision thanks to the fast capability to scan the fringes (less
pixels to be read at each sample). The broad-band mode will be
dedicated to imaging or limited spectral analysis.

The scientific observations will be interleaved by observations of
calibration stars with known (or very small) diameters in order to
estimate the value of the transfer function for the 6 visibilities and
the 3 closure phases. We plan to follow a very classical calibration
procedure, with a global fit of the transfer function estimates over
the night. The calibration stars will be chosen among the same
catalog as for AMBER. We do not expect any special difficulty to
find adequate calibration stars in the near-IR for the range of
baselines offered by the VLTI. The PIONIER data reduction software
will run in pseudo-real time on an offline machine in the VLTI control
room. Each new file will be processed and the new visibilities and the
closure phase will be added to the trend of the night. We believe this
is a critical tool to optimize the scientific return of the PIONIER
time. We also expect to compute the transfer function and to calibrate
the scientific data in pseudo-real time (say 15 min after the files
have been recorded), so that calibrated OIFITS data are available at
the end of the night.

Important to notice, the Auxiliary Telescopes of VLTI cannot be
relocated during the night. PIONIER will therefore make use of the
same 4T configuration during a full night. We expect that 2 nights of
observation at least are required to perform an image reconstruction.

\section{Data acquisition}
\label{sec:data}

\subsection{Interferometric signal description}

 \begin{figure}[t]
   \centering
   \includegraphics[width=0.4\textwidth]{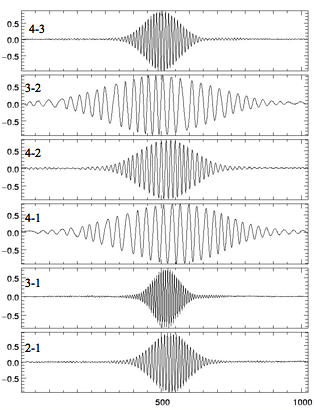} \\
   \includegraphics[width=0.35\textwidth]{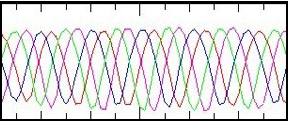}
  \caption{Top: Temporally encoded normalised interferograms (time: arbitrary units). One interferogram
    per baseline. Bottom: One ABCD output showing phase shifted
    interferograms almost in quadrature.}
   \label{fig:interferograms}
 \end{figure}

 If the interferometer was cophased, the IO combiner would allow the
 visibility and closure phases to be extracted from one single
 detector frame. Indeed the combining cell generates four outputs in
 almost phase quadrature. This combined with spectral dispersion,
 which provides an estimation of the group delay allows complex
 coherence to be retrieved on each baseline. But this IO combiner
 allows an easier, although less efficient, mean of recording coherent
 signal. By scanning fringes throughout the coherence envelope like in
 Fluor or VINCI we can recover the complex visibilities.  Obviously
 the presence of 4 beams requires to modulate the optical paths at non
 redundant velocities in order to generate the fringe
 signal. Typically three beams will be scanned at the respectives
 velocities of $v,-v,-2v$ with $v$ obtained after a compromise between
 sensitivity and the necessity to freeze as much as possible the
 atmospheric effects. Figure \ref{fig:interferograms} shows laboratory
 interferograms recorded on the PIONIER beam combiner (see Benisty et
 al. 2009 for the details). PIONIER data reduction will initially use the scanned
 interferograms information until a proper way to extract the coherent
 signal from the ABCD encoding is found.

The pairwise design allows to extract the simultaneous photometric
signal by a global fit to the coherent data. This simultaneity is the
key to obtain precisely calibrated interferograms.

\subsection{Data reduction}
The data reduction software of PIONIER will convert the raw data from
the detector into visibilities and phase measurements. Our strategy is
directly inspired from the FLUOR, VINCI and IOTA-3T experiments. In
these instruments, the scanning method associated with an estimate of
the visibility in the Fourier space (integration Power Spectral
Densities) provided sensitive and accurate measurements, without the
need of tricky or complicated internal calibration procedures.

The raw files are stored on the instrumental machine and follows the
ESO convention for VLTI data. The format has been only slightly
modifed to allow the recording of each individual read of the detector
pixels, and not only the final value (for instance using the DOUBLE
CORRELATED mode, the 2 read are stored individually). Starting from
these data, the PIONIER data reduction algorithm is composed of the
following steps:
\begin{itemize}
\item Cosmetic: We perform the very basic operation that are related
  to the detector in order to convert the raw reads into flux
  measurements. The consecutive non-destructive reads are subtracted
  and the average bias level is removed. More advance operation may be
  implemented latter on, once the behavior of the detector in real
  operation will be better known (optimal interpolation of the
  non-destructive reads, filtering...).
\item Photometric monitoring: For each baseline, the 4 ABCD outputs
  are summed together (taking into account the relative transmission)
  in order to obtain a measurement of the total photometry of the sum
  of the two considered telescopes. The result is a vector of 6
  measurements of the sum of two telescopes. This system is invertible
  and the individual photometries of each telescope can be
  recovered. This procedure was commonly used on the IOTA-3T beam
  combiner.
\item Computation of the clean fringe signals: The 24 raw signals are
  cleaned from the real-time photometric fluctuations of each
  telescope. The photometric continuum are removed and the fringe
  enveloppes are normalized. Then the 4 ABCD outputs of each baselines
  are co-added taking into account the relative phase shifts. The
  result is a single, clean, high-SNR interferogram per
  baseline. Thanks to the ABCD measurements, this interferogram is a
  complex quantity.
\item Measurement of the visibilities: The Power Spectral Density of
  each interferogram is stacked until the fringe peak appears above the
  noise. The power in the fringe peak is a un-biased measurement of the
  squared visibility.
\item Measurement of the phase closures: We expect to implement and
  compare several methods to estimate the phase closures.
\end{itemize}

\section{Conclusion: status and future work}
\label{sec:status}

The principle of PIONIER has been approved by STC at spring 2009. It
has the status of visitor instrument and competes with other VLTI
mainstream instruments for time allocation. Proposals are submitted
each semester for scientific evaluation by ESO’s OPC. First
commissioning and scientific lights are expected at the end of 2010
(Period 87). As of july 2010 PIONIER has been integrated in Grenoble
and is undergoing preliminary testing, four telescope internal fringes
have been recorded. The current version of PIONIER
only accomodates an H band combiner. The extension to the K band is
planned and will be developed in the coming year.

\acknowledgements

PIONIER is funded by the Universit\'e Joseph Fourier (Grenoble), the
INSU ``Programme National de Physique Stellaire'' and INSU ``Programme
National de Plan\'etologie'' and equipped with a detector provided by
W. Traub (JPL, Caltech). It is developed by the GRIL (now CRISTAL) instrumental team of
Laboratoire d'Astrophysique de `Grenoble (LAOG) in collaboration with
R. Millan-Gabet (Nextsci, Caltech) and W. Traub.

\bibliography{pionierSpie2010}   
\bibliographystyle{spiebib}   
\end{document}